# Ultrafast Spatiotemporal Control of 3D Angular Momentum Vectors Drives the Spin and Orbital Precession of Light

Zhe Zhao[1,†], Jinfeng Liu[1,2,3], Yize Liang[1,2,3], Xiaokang Lian[1,2,3], and Jian Wang[1,2,3*]

[1]Wuhan National Laboratory for Optoelectronics and School of Optical and Electronic Information, Huazhong University of Science and Technology, Wuhan 430074, Hubei, China

[2]Hubei Optical Fundamental Research Center, Wuhan 430074, Hubei, China

[3]Optics Valley Laboratory, Wuhan, 430074, Hubei, China

[†]Present address: Department of Physics, University of Muenster, Muenster 48159, Germany

*Corresponding author: jwang@hust.edu.cn

**The spin and orbital angular momenta (AM) of light are three-dimensional (3D) vectors associated with two well-established forms of handedness: polarization and helical phasefront, respectively. Recently, the ultrafast control of optical AM dynamics in one or two dimensions has attracted increasing interest, driving an AM vector to change in magnitude or sign, or undergo tilting within a fixed plane. In three dimensions, however, an AM vector can evolve over a curved surface or within a structured volume, opening up largely unexplored possibilities for new AM dynamics and forms of optical handedness. Here, we demonstrate optical AM precession up to 493 GHz, a fundamental mode of 3D AM-vector evolution, and control its conical trajectory on picosecond timescales. Our findings establish AM precession as an additional form of dynamic handedness in optical beams, and provide a versatile approach to synthesizing and characterizing structured optical beams with increasingly sophisticated 3D AM dynamics.**

## Introduction

Optical beams carry angular momentum (AM) in two forms: spin AM (SAM), associated with circular polarization, and orbital AM (OAM), associated with helical phasefronts (1–3). Both are vectors in three dimensions (3D), each with a well-known handedness: that of the circular polarization for SAM and that of the helical phasefront for OAM. In a paraxial monochromatic beam, they point along the propagation direction, with projections $s\hbar$ and $\ell\hbar$ per photon that label the spin polarization and the orbital helical phasefront of the beam (4, 5). The AM of optical beams underpins optical manipulation, super-resolution imaging, high-capacity communications, and quantum information processing (6–9), and spatiotemporal manipulation of optical beams has turned it into a designable quantity (10–13). Spatiotemporal optical beams (STOBs) can carry both SAM and OAM transverse to the propagation (14–19), such that the AM vector can be given a controllable but static orientation. Moreover, STOBs can also have time-variant SAM and OAM along a fixed axis or plane (Fig. 1a). For example, SAM and OAM can switch their sign or change their value along a fixed axis in STOBs, giving optical beams a self-torque (20–28). OAM can also change its value and orientation in a fixed plane (Fig. 1a) by exploiting coordinate transformations (19) or by intersecting multiple STOBs (18), which enables more sophisticated structures such as toroidal vortices close the vortex line into a ring (29). It would thus be desirable to manipulate the AM vector in 3D, such as a time-variant AM within a curved surface or a structured volume, which might enable more advanced dynamics that cannot be observed in a line or a plane.

In matter, one of the elementary dynamics of an AM vector is precession: its direction precesses around a fixed axis while its magnitude remains constant, which makes it a suitable example for synthesizing and characterizing STOBs whose spin and orbital AM vectors follow a continuous trajectory in 3D. The precession of electron spins (Fig. 1c) in a magnetic field underlies magnetic resonance (30–32), and the OAM of electron vortex beams precesses in the same way (33). These precessions are driven by the torque that the field exerts on a magnetic moment. However, photons have no magnetic moment, and no external torque acts on an optical beam in free space, whose total AM is conserved. Therefore, a possible approach to achieving a precessing AM vector of optical beams is to build it into the spatiotemporal field itself.

Fields whose intensity and phase profiles rotate about the propagation axis have been synthesized by correlating frequency with OAM, as in optical springs and vortex frequency combs (34–41). In these fields the sense of rotation is set by the sign of the frequency–OAM correlation and can be reversed (40), but only the longitudinal OAM along the propagation axis has been synthesized and characterized. Theory has described a precession and nutation of OAM (42) and predicts that the SAM of two intersecting optical beams of different frequencies processes at their frequency difference (43). To our knowledge, however, manipulating and characterizing either optical SAM or OAM along a continuous trajectory in 3D has remained out of reach in experiment.

Here we manipulate STOBs whose AM vector precesses in a programmable manner on picosecond timescales and measure the SAM, OAM and total AM vectors at each time slice. The sense of the AM precession is an additional handedness, independent of the handedness of the polarization and of the helical phasefront (Fig. 1b). We achieve this with five equally spaced spectral subregions of an optical pulse, each carrying a superposition of Laguerre–Gaussian ($LG_{\ell,p}$) modes containing a different $\ell$ value but multiple $p$ values. Each time slice of the synthesized STOBs is, to a good approximation, a rotated copy of the first, so its AM vector precesses about the propagation axis at the frequency spacing between neighboring spectral subregions. Characterizing the complete field of each time slice, we observe AM precession at 131.5–493.0 GHz, with total AM from $-3\hbar$ to $+3\hbar$ per photon, precession angles of 0 to 0.650 mrad, and both precession handedness, for the spin, orbital and total AM vectors. We believe that our observations could be useful for synthesizing and characterizing structured optical beams with increasingly sophisticated 3D AM dynamics.

## Results

### Manipulating the AM vector with frequency–OAM correlation

We use the elementary AM precession on a cone as an illustrative example of AM dynamics in 3D. Figures 1c and 1d compare two ways in which an AM vector can precess. An electron in a magnetic field $\mathbf{B}$ experiences a torque on its magnetic moment, and its AM precesses about $\mathbf{B}$ at the Larmor frequency (Fig. 1c) (30, 31). An optical pulse in free space experiences no external torque, and its total AM is conserved. A precessing AM vector of optical beams could instead be obtained by redistributing the AM among the time slices of the pulse through spatiotemporal synthesis (10–13).

Mathematically, an STOB exhibiting AM precession could be constructed as follows: a Laguerre–Gaussian beam ($LG_{\bar{\ell},\bar{p}}$) with polarization $\hat{\mathbf{e}}$ is first tilted by a small angle $\alpha_p$ away from the $z$ axis, and then the direction of the tilt is made to precess about $z$ at the angular frequency $\omega_p$ with $|\omega_p| = 2\pi f_p$, where $\bar{\ell}$ and $\bar{p}$ are the azimuthal and radial mode indices, and $f_p$ is the precession frequency. At the synthesis plane $z = 0$ the field reads (Supplementary Note 3, Eq. (S30))

$$\mathbf{E}_{STOB}(r,\varphi,t) = s(t)\, u(r) e^{-j\bar{\ell}(\varphi-\omega_p t)} \cdot e^{-jk_0\alpha_p r\cdot\cos(\varphi-\omega_p t)} \cdot e^{j\omega_0 t}\, \hat{\mathbf{e}}\,, \tag{1}$$

where $(r,\varphi)$ are polar coordinates, $u(r)e^{-j\bar{\ell}\varphi}$ is the LG mode profile, $s(t)$ is the temporal envelope, $k_0 = \omega_0/c$ and we use the convention $e^{-i\omega t}$ (j = −i in our equations), in which a phase $e^{i\ell\varphi}$ carries an OAM of $\ell\hbar$ per photon. Expanding the rotating tilt with the Jacobi–Anger identity gives its spectral content (Supplementary Note 3, Eq. (S32))

$$\mathbf{E}_{STOB}(r,\varphi,t) = \sum\sum_{\ell,p} C_{\ell,p} LG_{\ell,p}(r,\varphi) \cdot s(t) \cdot \mathrm{e}^{j(\omega_0+\ell\omega_p)t}\hat{\mathbf{e}}. \tag{2}$$

where $C_{\ell,p}$ is the complex coefficient (Methods), $LG_{\ell,p}(r,\varphi)$ is the electric field of an $LG_{\ell,p}$ mode (see Eq. (S14) in Supplementary Note 2). The field is a comb of spectral lines spaced by $\omega_p$, in which line $\ell$ carries the OAM $\ell$. This frequency–OAM correlation, $\omega_\ell - \omega_0 = \ell\omega_p$, is the rotational Doppler relation of a field that rotates rigidly at $\omega_p$ (44, 45). Conversely, imposing this correlation on a comb whose lines carry the LG-mode content of a tilted $LG_{\bar{\ell},\bar{p}}$ beam produces a pulse in which every time slice is a copy of the first one rotated about $z$ by the angle $\omega_p t$ (Methods). Every AM vector of a slice, namely its spin AM $\mathbf{S}$, its orbital AM $\mathbf{L}$ and its total AM $\mathbf{J}$, therefore precesses about $z$ with the same angular velocity

$$\frac{d\mathbf{J}}{dt} = \omega_p\, \hat{\mathbf{z}} \times \mathbf{J}. \tag{3}$$

Equation (3) has the form of the equation of motion of a precessing magnetic moment carried by an electron, which thus could be considered as a mathematical, but not a dynamic, analogy of the electronic Larmor precession whose origin is different. Equation (3) describes how the AM is distributed over the successive time slices of one coherent pulse, a texture written by the spectral correlation, and $t$ is the time at which a slice reaches the plane of observation. An observer at a fixed plane therefore sees an AM vector whose direction precesses at $\omega_p$. The precession frequency is the comb spacing, the cone angle is set by $\alpha_p$, the handedness by the sign of the correlation and the magnitude by $\bar{\ell}$ and the polarization. Equations (1)–(3) encode the handedness of the AM precession in the sign of $\omega_p$: $\omega_p > 0$ for the correlation of Eq. (2) gives anticlockwise precession seen from $+z$, and $\omega_p < 0$, the mirrored correlation in which the line at $\omega_0 - \ell|\omega_p|$ carries the OAM $\ell$, gives clockwise precession. Rego et al. called the rate of change of the OAM of successive time slices a self-torque (21). In that sense, $d\mathbf{J}/dt$ is a self-torque that is perpendicular to $\mathbf{J}$ and rotates with it, rather than parallel to the propagation axis. We notice that it implies no exchange of AM with another system, but a redistribution of AM within the optical pulse in time.

## Transforming simple LG beams to STOBs with AM precession

We now consider the synthesis of STOBs exhibiting optical SAM and OAM precession. Figure 2 illustrates how the correlation between frequency and OAM is used, in three cases and two transformation steps.

In the first transformation step, we introduce an angular tilt to the average Poynting vector and AM (Figs. 2b and 2d). A conventional optical $LG_{\bar{\ell},\bar{p}}$ beam has the electric field (3):

$$\mathbf{E}_{STOB}(r,\varphi,z,t) = LG_{\bar{\ell},\bar{p}}(r,\varphi,z) \cdot s(t) \cdot e^{j\omega_0 t} \cdot \hat{\mathbf{e}}. \tag{4}$$

The $LG_{\bar{\ell},\bar{p}}$ beam features a helical phasefront that twists around its center over time (Figs. 2a and 2c). The 'twisting' motion is induced by the spatially homogeneous phase term $e^{j\omega_0 t}$ in Eq. (4), resulting in AM whose direction remains parallel to the positive *z*-axis. Such a structured beam can maintain its intensity and phase profiles while its propagation direction is tilted from the *z*-axis by an angle of $\alpha_p$. This is achieved by combining multiple modes from a complete modal basis set at a single center frequency, with each mode being accompanied by a judiciously selected complex coefficient (46, 47). As illustrated in Figs. 2b and 2d, the constructive or destructive interference of several $LG_{\ell,p}$ modes at a single center frequency results in an optical beam with intensity and phase profiles identical to those of an $LG_{\bar{\ell},\bar{p}}$ beam, whose beam propagation direction is tilted from the original *z*-axis by a certain angle. The electric field of the $LG_{\bar{\ell},\bar{p}}$ beam can be written as $\sum\sum_{\ell,p} C_{\ell,p} LG_{\ell,p}(r,\varphi,z) \cdot s(t) \cdot e^{j\omega_0 t}\mathbf{\hat{e}}$, where $C_{\ell,p}$ denotes the complex coefficient of each $LG_{\ell,p}$ mode. The static phase delays ($\Delta\varphi$) between different $LG_{\ell,p}$ modes are defined as the phase differences of the complex coefficients $C_{\ell,p}$, written as $\Delta\varphi = \angle C_{\ell_1,p_1} - \angle C_{\ell_2,p_2}$, where the operator $\angle$ calculates the phase of a complex value. Notably, $\Delta\varphi$ is time-independent. Therefore, at different times $t_1$ and $t_2$, the coherent superposition of all modes leads to a static term $\sum\sum_{\ell,p} C_{\ell,p} LG_{\ell,p}(r,\varphi,z)$, multiplied by a time-dependent factor of $s(t) \cdot e^{j\omega_0 t}\mathbf{\hat{e}}$ that governs the temporal evolution of the field amplitude and global phase. As shown in Fig. 2b, the spatial phase and intensity profiles remain almost unchanged over time, with the only differences arising from a global phase shift of $\omega_0 \Delta t$ and an intensity variation due to the envelope of $s(t)$.

In the second transformation step, we introduce the dynamic optical SAM and OAM precession. For the above superposition, the constructive or destructive interference of multiple modes at a single center frequency experiences time-invariant phase delays. Therefore, it produces an $LG_{\bar{\ell},\bar{p}}$ beam with its intensity and phase profiles appearing static at a given distance. To achieve an additional dynamic precession around the *z*-axis, appropriate time-variant relative phase delays among the modes could be applied. We achieve this by applying a time-variant relative phase delay of $\Delta\varphi = \omega_p t$ between neighboring $LG_{\ell,p}$ modes (*i.e.*, $\Delta\ell = 1$) carried by multiple frequency subregions or comb lines. Each frequency component contains several $LG_{\ell,p}$ modes with a different $\ell$ value and multiple $p$ values (Fig. 2e). Here, $\omega_\ell = \omega_0 + \ell\omega_p$ denotes the center angular frequency of each $LG_{\ell,p}$ mode, where $\omega_p$ is the angular precession frequency as well as the frequency spacing between adjacent frequency components, and $\omega_0$ is the center angular frequency of the tilted $LG_{\bar{\ell},\bar{p}}$ beam without the dynamic precession motion. Prior studies have shown that a relative phase delay of $\Delta\varphi$ between neighboring $LG_{\ell,p}$ modes at a single frequency can rotate the azimuthal position of the resultant beam by a static angle $\Delta\theta = \Delta\varphi$, and that a time-varying delay $\Delta\varphi = \Delta\omega t$ rotates it continuously (35, 36). In this paper, the time-variant phase delays induce dynamic constructive and destructive interferences, synthesizing an $LG_{\bar{\ell},\bar{p}}$ beam with an additional phase tilt whose direction rotates in time. This STOB not only carries AM but also exhibits AM precession around the *z*-axis (Fig. 2f).

**Synthesis and spatiotemporal reconstruction**

Here, we detail the methods for synthesizing STOBs exhibiting AM precession. As illustrative examples, we first show two simple cases: one STOB carrying only OAM with $\bar{\ell} = +1$, and another carrying only SAM with $s = +1$. We follow the transformation schemes (Fig. 2) to synthesize these two STOBs. Specifically, we first obtain the complex coefficients $C_{\ell,p}$ for each $LG_{\ell,p}$ mode carried by: (a) an $LG_{1,0}$ beam with a linear polarization along the *x*-axis, and (b) a Gaussian beam with LCP, both at a single frequency and tilted by an angle of $\alpha_p$ from the *z*-axis. We then correlate these $LG_{\ell,p}$ modes with complex coefficients to different frequency components based on the OAM value $\ell$.

We now address the experimental synthesis of these STOBs with the above-described correlations between frequency and OAM. The basic concept (illustrated in Fig. 3a) involves ultrafast pulse shaping (13), 2D complex spatial beam modulation (48), and grating- and SLM-based beam splitting and combining to introduce the frequency–OAM correlation (24, 36). We start from a custom-built fiber-based short-pulsed laser as the coherent light source, in which all the frequency components carry the same $LG_{0,0}$ (*i.e.*, Gaussian) mode on the *x* polarization. We then use a dispersion box, consisting mainly of a grating and a spatial light modulator (SLM1), to shape the frequency spectrum and introduce an amplified dispersion along the *y*-axis. In detail, we select five equally spaced spectral subregions and use SLM1 to encode each subregion with a linearly varying phase along the *y*-axis whose slope corresponds to a specific wavevector $k_y$. These phase gradients along the *y*-axis project each spectral subregion onto a specific area of a second SLM2, where each spectral subregion with a central angular frequency of $\omega_0 + \ell\omega_p$ is modulated with a 2D complex spatial amplitude and phase profile of $\sum_p C_{\ell,p} LG_{\ell,p}(r, \varphi, z)$ (48). A third SLM3, acting as a combination of five grating patterns with specific values of $k_y$, recombines all the frequency and mode components and directs them into the same direction, which synthesizes the STOB carrying OAM. Finally, depending on the SAM value of the synthesized STOBs, a quarter-wave plate with its fast axis at 0°, +45° or −45° sets the polarization vector $\hat{\mathbf{e}}$ to linear *x* polarization, RCP or LCP, respectively. Experimental details are given in Methods and Supplementary Note 5.

We experimentally synthesize these two STOBs through the coherent combination of five spectral subregions from the input optical pulse. Each spectral subregion consists of multiple $LG_{\ell,p}$ modes and can be written as $\sum_{\ell,p} C_{\ell,p} LG_{\ell,p}(x, y, z)$, where $\ell$ and $p$ range from $\bar{\ell} - 2$ to $\bar{\ell} + 2$ and 0 to +4, respectively. Figures 3b and 3c show the power and phase distributions of each $LG_{\ell,p}$ mode for synthesizing the two STOBs. These results indicate that more than 98.8% of the power is concentrated in modes with $\ell$ ranging from $\bar{\ell} - 2$ to $\bar{\ell} + 2$, so that five spectral subregions suffice to synthesize STOBs with a theoretical OAM mode purity above 98.8%. Figures 3d to 3g show the measured spatiotemporal evolutions of the STOBs' amplitude/intensity, phase, and polarization profiles. To characterize the spatiotemporal properties of the synthesized STOBs, we use an off-axis interference measurement (16, 36, 49) to resolve the spatial, temporal and polarization profiles (Methods). We first present the iso-intensity (10% of the maximum intensity) profiles of the synthesized STOB exhibiting AM precession over a 10-ps time window at a specific distance $z = 0$ (Figs. 3d and 3f). The STOB exhibiting only OAM precession (Fig. 3d) has an iso-intensity profile (red shell in the left-bottom coordinate) with a 'vortex hollow' cylindrical shape. The inserted slices are the corresponding phase profiles in the *x*-*y* plane at different times, presenting a 'rotation' motion around the center hollow in time. The STOB's polarization remains along the *x* direction (*i.e.*, no SAM), whose electric field has an *x*-component $E_x$ varying in time, and *y*-component $E_y \approx 0$ (see the right-top coordinate). Figure 3e shows further details of the temporal evolutions of the STOB's amplitude and phase profiles from 2.53 ps to 7.87 ps. Its amplitude profile (first row) has a ring-like shape, a feature of non-zero OAM value, and its phase profile (second row) is the sum of a helical phase and a gradient phase, with its slope direction precessing around the *z*-axis in time (third row). Compared with the helical phase profiles in Figs. 2b and 2e, the subtracted helical phase profile in Fig. 3e contains an additional lens-like phase term of $\beta(x^2 + y^2)$, where $\beta$ is a constant for each STOB. This $\beta r^2$ phase term is subtracted to compensate for diffraction effects of STOBs due to free-space propagation. In comparison, the STOB exhibiting only SAM precession (Fig. 3f) has an iso-intensity profile of 'Gaussian solid' cylindrical shape and circular polarization. Figure 3g shows the temporal evolutions of its amplitude, polarization and phase profiles. It has a dot-like amplitude shape (first row), a polarization of almost LCP ($s \approx +1$, second row), and only a gradient phase profile whose slope direction changes in time.

**Characterization of the precessing AM vector: trajectory, angular impulse and handedness**

Next, we characterize the dynamic AM precessions of the synthesized STOBs by analyzing the 3D AM vector in each time slice of STOBs. The theory and measurement of the AM vector are detailed in Methods and Supplementary Note 4. Figures 4a and 4d show the measured total AM vector $\mathbf{J}$ of successive time slices of the two elementary STOBs in Figs. 3d–g. Its transverse part, about $1.6 \times 10^{-4}\ \hbar$ per photon for the OAM-only STOB, precesses anticlockwise about $z$ and completes one turn in about 4.1 ps, close to the design period of 4.06 ps. That is, the AM precesses at about 243.9 GHz. Over the central 5.3 ps, where the STOBs carry more than 97% of their energy, the power in the targeted OAM component stays above 73% (OAM-only STOB) and 76% (SAM-only STOB) and peaks at 91% and 95% (Figs. 4b and 4e). Outside the central 5.3 ps, the power in the targeted OAM component decreases, falling below 10% at the edges. This could be explained by the increase in the measurement error (*i.e.*, noise) as the power of the STOBs decreases, leading to a reduction in the spatial mode purity. Because the mode purity decreases while the power noise remains at a similar level, the error becomes larger outside the central 5.3 ps. We believe that the fluctuation in the mode purity and the total AM could be reduced by (i) increasing the power of the synthesized STOBs, and (ii) using a semiconductor camera with a higher sensitivity for detection.

Furthermore, we measure the accumulated changes of $\mathbf{J}$ along its azimuthal direction to quantify the dynamics of the transverse AM vector $\mathbf{J}_\perp$. Integrating the azimuthal and radial components of $\mathbf{J}$ over time gives the accumulated transverse AM, labelled the azimuthal and radial angular impulses in Figs. 4c and 4f. The angular impulse along a direction $\hat{\mathbf{d}}$ is thus defined as

$$\Lambda_d(t) = \int_{t_0}^{t_1} \frac{d\mathbf{J}}{dt} \cdot \hat{\mathbf{d}}\, dt. \tag{5}$$

The angular impulse $\Lambda_\theta(t)$ along the azimuthal direction comes with the precession of AM around the $z$-axis, while the angular impulse $\Lambda_r(t)$ along the radial direction describes an offset of the AM along the radial direction. We refer to the second motion as nutation of the optical AM (42), analogous to the nutation of electrons in magnetic fields (32). The measured angular impulse $\Lambda_\theta(t)$ appears to be linearly dependent on time in the central range (2.53 ps to 7.87 ps), with slopes of 0.303 and 0.395 $\hbar \cdot \mathrm{ns}^{-1}$ for the OAM-only and the SAM-only STOBs, respectively (Figs. 4c and 4f). This agrees with our expectation of a constant rate of change of the total AM $\mathbf{J}$. Meanwhile, the measured angular impulse along the radial direction, $\Lambda_r(t)$, fluctuates around zero, with an absolute amplitude of less than $6 \times 10^{-4}\ \hbar$. A larger $\Lambda_r(t)$ would lead to faster nutation of AM and stronger fluctuations of AM along the radial direction. Given that $\Lambda_r(t)$ is much smaller than $\Lambda_\theta(t)$, the AM of the STOBs mainly precesses about the z-axis rather than nutating along the radial direction. The measurement error in the angular impulse shown in Figs. 4c and 4f (as well as in Fig. 5) can be understood mathematically using Eq. (5). It shows that the measurement errors in $\Lambda_\theta(t)$ and $\Lambda_r(t)$ depend linearly on the measurement error in the change of the total AM $\mathbf{J}$.

Therefore, a larger error in the total AM (see Figs. 4b and 4e) leads to a larger error in the angular impulse outside the central range. Similarly, the fluctuation observed in the angular impulse in Figs. 4 and 5 could also be reduced through minimizing the measurement error of the total AM **J**, as described in (i) and (ii) in the previous paragraph.

In addition, the AM precession (OAM **L**, SAM **S** and total AM **J**) can have the same handedness for STOBs carrying different AM values. From Figs. 4a and 4d, we note that the AM precession exhibits the same anticlockwise handedness for both OAM-only and SAM-only STOBs. Both AM vectors **J** precess anticlockwise seen from $+z$. We show the measured AM precession for three more cases of STOBs, which carry both OAM and SAM with targeted values of $(\bar{\ell}, s) = (+1, +1)$, $(-1, +1)$ and $(-1, -1)$ in Figs. 4g–i. The non-zero OAM **L**, SAM **S** and total AM **J** all precess anticlockwise seen from $+z$ at about 246.5 GHz (Supplementary Movie 1 for animation). These results show that the handedness of the polarization and of the helical phasefront can be changed while the handedness of the AM precession stays the same.

**Programmable AM precession: AM value, angle, speed and handedness**

STOBs with AM precession can be synthesized and programmed by selecting the correlation between the frequency and OAM components. Here, we investigate the programmability of our experimental setup for synthesizing STOBs exhibiting AM precession. This involves varying their AM (both OAM and SAM) values, as well as adjusting the precession angle, speed, and handedness (Fig. 5, Supplementary Figs. S6–S9 and Supplementary Movies 2–5).

To begin with, we vary the AM value of the synthesized STOBs. Figures 5a and 5d show the angular impulses along the azimuthal direction for STOBs exhibiting an anticlockwise OAM precession with an OAM value from $\bar{\ell} = -3$ to +3 (Supplementary Movie 2 for animation). The results indicate that the angular impulse component $\Lambda_\theta(t)$ has the same sign as $\bar{\ell}$, and its absolute value increases faster over time when the absolute OAM value $\bar{\ell}$ increases. This can be explained by analyzing the relationship between $\Lambda_\theta(t)$ and the AM change rate $d\mathbf{J}/dt$, using the formula $\Lambda_\theta(t) = \int_{t_0}^{t_1} d\mathbf{J}/dt \cdot r \cdot \hat{\boldsymbol{\theta}} dt$, which is derived from Eq. (5). When STOBs exhibit an anticlockwise OAM precession with a positive (or negative) OAM value $\bar{\ell}$, there will be a positive (or negative) OAM change along the positive azimuthal $\theta$ direction. Mathematically, $\Lambda_\theta(t) \propto d\mathbf{J}/dt \cdot \hat{\boldsymbol{\theta}}$ is positive (or negative) for a positive (or negative) OAM value $\bar{\ell}$. Moreover, we emphasize a special case when an STOB carries a zero OAM value $\bar{\ell} = 0$ (turquoise line labeled $LG_{0,0}$). Although the mean wavevector still exhibits dynamic precession around the $z$-axis, there is no theoretical AM precession. This is because when the synthesized STOB carries a zero AM value, there is only a linear momentum change but no AM change over time. Finally, the effects of OAM and SAM precession on the angular impulse tend to be equivalent (Figs. S9c and S9d in Supplementary Note 9). By comparing the STOBs exhibiting only OAM precession (Figs. 5a and 5d) and those exhibiting both OAM and SAM precession (Figs. S9c and S9d), we find that the angular impulse component $\Lambda_\theta(t)$ changes over time with a similar slope, provided that the total AM has the same value. This equivalence arises from the linear dependence $\Lambda_\theta(t) \propto d\mathbf{J}/dt \cdot \hat{\boldsymbol{\theta}} \propto (\bar{\ell} + s)$, when the STOBs have the same precession of the mean wavevector. Hence, $\Lambda_\theta(t)$ depends on the total AM rather than on its distribution between OAM and SAM.

Next, the dynamic precession motion of the synthesized STOBs can also be controlled. Figures 5b and 5e show the angular impulse along the azimuthal direction for STOBs exhibiting an anticlockwise OAM precession with a target precession angle ranging from $\alpha_p = 0$ to 0.650 mrad (Supplementary Movie 3 for animation). For reference, we include a case with a zero precession angle, achieved by generating an $LG_{1,0}$ beam on a single $LG_{1,0}$ mode carried by a single spectral subregion of the short pulse. The angular impulse component $\Lambda_\theta(t)$ changes at a higher rate over time as the precession angle increases. This is because a larger precession angle tilts the AM further from the z-axis, which increases the rate of change of the AM component in the transverse $(r, \theta)$ plane. Consequently, the resulting angular impulse component $\Lambda_\theta(t)$ changes at a higher rate over time. Additionally, we vary the precession frequency from $f_p$ = 131.5 GHz to 493.0 GHz (Figs. 5c and 5f, and Supplementary Movie 4 for animation). The tunability of precession frequency is achieved by increasing the frequency gap $\Delta f = f_p$ between the neighboring spectral subregions. This is because the precession is driven by the time-variant phase delay $\Delta\varphi = \omega_p t = 2\pi f_p t$ between the neighboring $LG_{\ell,p}$ modes (*i.e.*, $\Delta\ell = 1$). Therefore, the constructive and destructive interference of all the modes results in a dynamic precession with a frequency of $f_p$. Last, we flip the STOBs' precession direction from the anticlockwise trajectory to the clockwise one (Figs. S9a and S9b in Supplementary Note 9, and Supplementary Movie 5 for animation). By contrast, for a particle in a magnetic field the sense of precession is fixed by the sign of its gyromagnetic ratio and can be reversed by changing the particle, for example from negative to positive charge (30) (Fig. 1c). For the STOBs, we achieve a flip of the AM precession direction by switching between two mirror-symmetric frequency–OAM correlations (Fig. 5g, and Figs. S9a and S9b in Supplementary Note 9). For instance, instead of correlating the $\ell$-th frequency subregion at the center angular frequency $\omega_0 + \ell\omega_p$ with a spatial pattern of $\sum_p C_{\ell,p} LG_{\ell,p}(x, y, z)$, we correlate the frequency subregion at $\omega_0 - \ell\omega_p$ with the same spatial pattern.

Finally, we show the predicted change rate of the angular impulse $\mathrm{d}\Lambda_\theta(t)/dt$ versus the AM value, precession angle, and precession frequency in broader ranges (Figs. 5g to 5i). The measurements in the limited ranges of Figs. 5a–5c are included to partially verify the accuracy of our predictions. Both the theoretical predictions and experimental measurements indicate that $\mathrm{d}\Lambda_\theta(t)/dt$ is linearly dependent on the total AM, $\sin\alpha_p$ and $f_p$, as expected from Eqs. (3) and (5). $\mathrm{d}\Lambda_\theta(t)/dt$ appears linearly dependent on the precession angle $\alpha_p$, given the approximation that $\sin\alpha_p \approx \alpha_p$ in the range from 0 to 100 mrad. This dependence can be understood by considering the AM change rate: when the STOBs have (i) a larger AM value, (ii) a larger precession angle, or (iii) a higher precession frequency, they experience a higher AM change rate, thus increasing the change rate of $\Lambda_\theta(t) \propto d\mathbf{J}/d\mathrm{t} \cdot \hat{\boldsymbol{\theta}}$.

In summary, the AM value, precession angle, speed and handedness of the synthesized STOBs can be independently controlled by tailoring the frequency–OAM correlation. Across all experimental configurations, the measured slopes follow the predicted dependence on the total AM, the cone angle and the handedness, with a median magnitude reaching approximately 74% of the targeted value. The deviations couldbe traced mainly to three factors: the measured longitudinal AM $|J_z|$, the cone angle of $\mathbf{J}$, and the precession speed of $\mathbf{J}$ deviate from their target values by 25%, 43% and 24%, respectively. We believe these deviations could be mainly due to the imperfect interference among multiple frequency and spatial mode components, arising from power and phase errors (mismatch from Figs. 3b and 3c). These errors could distort the helical phasefronts and alter the tilt angle, thereby leading to corresponding deviations in the resulting angular impulse. Adding a feedback loop to compensate for these power and phase errors (50, 51) could potentially reduce the mismatch between the measured and target values.

## Discussion

We have experimentally demonstrated that the AM vector of optical beams can be manipulated on a cone in 3D on picosecond timescales in a programmable manner. A frequency–OAM correlation across five comb lines makes the spin, orbital and total AM vectors of successive time slices precess about the propagation axis at the comb spacing, and the magnitude, cone angle, speed and handedness of the precession are set independently by the design of the correlation. Measuring the AM vectors of the time slices from the reconstructed field, we observe this precession for spin, orbital and combined AM. The sense of this precession could serve as an additional handedness of optical beams, besides those of the polarization and the helical phasefront. Our findings highlight the potential capabilities of utilizing the correlation of optical beams' multiple degrees of freedom (DoFs) to synthesize and characterize structured optical beams with more sophisticated AM dynamics in 3D on ultrafast timescales.

We note that our experiments primarily focus on synthesizing STOBs exhibiting optical AM precessions with range-limited dynamic properties, including a sub-mrad precession angle, a sub-THz precession frequency, and a sub-$\hbar \cdot ns^{-1}$ change rate of angular impulse. However, we believe that it might be possible to extend these dynamic ranges through the following approaches:

(i) *AM value*: the AM value is limited in our setup for the following reasons: (a) to filter out undesired orders from the SLM3 (Fig. 3a) when coherently combining multiple modes and frequencies, we insert a spatial filter with a limited aperture in the optical path; and (b) because optical beams with larger OAM values diffract faster than those with lower OAM values (3), the beams would have a larger beam size at the filter aperture and thus be partially obstructed. Therefore, the limited aperture worsens the coherent interference of higher-order modes, limiting the performance in synthesizing STOBs exhibiting AM precession with a high AM value. One potential approach to synthesizing an STOB with a high OAM value (*e.g.*, $\ell_2 = 100$) could involve directly projecting the STOBs ($\ell_0 = +1$, $s_0 = 0$) in our current experiment onto an additional SLM with a helical phase pattern of $e^{-j\ell_1\varphi}$, where $\ell_1 = \ell_2 - (\ell_0 + s_0)$.

(ii) *Precession angle*: the largest precession angle in our experiments is 0.650 mrad. Increasing the precession angle increases the tilt of the STOBs' phasefronts and splits their spatial mode components and their frequency spectra into a large number of $LG_{\ell,p}$ modes with larger $\ell$ and $p$ values, and a greater number of frequency spectral subregions (3). Achieving large numbers in our current experimental setup is challenging, but there have been reports of synthesizing STOBs with extensive modes and frequency subregions or comb lines, using techniques such as: (a) multi-plane light converters to generate ~ 210 $LG_{\ell,p}$ modes (52) and ~ 90 frequency components with each carrying different modes (53); (b) Fourier space-time shapers for ultrabroadband pulses that carry a wide range of OAM modes with OAM values up to 80 (37); and (c) integrated vortex soliton combs supporting 50 frequency components, each one carrying an OAM mode with a distinct value (38, 39). We believe these techniques could be utilized to enhance the experimental feasibility of STOBs precessing at large angles and carrying high AM values with high mode purity.

(iii) *Precession speed*: our fiber-based short-pulsed laser has a 3-dB bandwidth of ~ 1.03 THz, and our spectrum-related manipulation devices have a resolution of ~ 0.411 GHz/μm × 8 μm = 3.29 GHz, given an SLM pixel size of 8 μm and a grating with an effective dispersion of 0.411 GHz/μm. Potential methods to broaden the precession speed range from several MHz to THz could be: (a) using a pulse source with a 3-dB bandwidth of tens of THz to raise the upper bound; and (b) employing dispersive components with a higher effective dispersion, such as cascaded gratings (54).

We also note that the synthesized STOBs, exhibiting optical SAM and OAM precessions, serve as preliminary examples for synthesizing STOBs with ultrafast AM dynamics in 3D. However, we believe it might be possible to synthesize STOBs with more complex and sophisticated motions along different trajectories. One direct idea is to reproduce the reported optical spin and orbital Hall effects (4, 55–57) through utilizing the correlation of optical beams' multiple DoFs. By enhancing this correlation, it might also be possible to achieve strong optical spin and orbital Hall effects, beyond the previously reported approaches, such as using metasurfaces with significant phase discontinuities to achieve strong spin–orbit interactions (55). Furthermore, if STOBs exhibit two distinctive correlations of their multiple DoFs, it might be possible to control their motions with two correlated trajectories simultaneously.

## Methods

### Experimental details for the STOB synthesis

We use a custom-built fiber-based short-pulsed laser as our coherent light source. Its center wavelength is set at around 1565 nm, the pulse width is around 299 fs, and the repetition rate is around 29 MHz. The short pulse is first amplified to around 10.37 dBm using a short-pulsed amplifier and then transmitted through a single mode fiber to a free-space collimator. The pulse, with all its frequency components in the $LG_{0,0}$ (*i.e.*, Gaussian) mode, propagates in free space for 20 cm before being split into two copies by a beam splitter. One copy is used for the synthesis of STOBs, and the other serves as a reference for reconstructing the spatiotemporal profiles of the synthesized STOBs (Fig. S1 in Supplementary Note 5).

A blazed grating with 600 grooves/mm is used to disperse the synthesized copy along the $x$-axis. To increase the spectral resolution in our setup, we use a 4f imaging system combined with an SLM1 to amplify the dispersion. The 4f imaging system, consisting of two cylindrical lenses ($f_1$ = 200 mm and $f_2$ = 400 mm), first projects the spread wavelength spectrum onto SLM1. We select five equally spaced subregions, each with a bandwidth of $f_p$, and a frequency gap $\Delta f = f_p$ between neighboring subregions, where $f_p$ is the precession frequency. Each subregion is modulated with a gradient phase along the $y$-axis, encoding each subregion with a specific tilt angle along the $y$-axis. The dispersion is amplified by up to about 4.7 times from 0.68 mrad/nm to 3.20 mrad/nm. All the spectral subregions are then sent back through the pair of cylindrical lenses and recombined to the same spot on the blazed grating, where the dispersion along the $x$-axis is cancelled, leaving only the dispersion along the $y$-axis.

A second 4f imaging system is built to: (i) introduce the designed correlation between frequencies and spatial modes; and (ii) project all the frequency components from one spot on the dispersive blazed grating in the object plane to another single spot in the imaging plane. The first lens with $f_1$ = 250 mm in this $4f$ system projects the five spectral subregions into five areas on SLM2. SLM2 is modulated with a lens phase profile, representing a physical lens with $f$ = 250 mm at the position from $f_1 + f_2$ = 500 mm to the first lens. Additionally, each area on SLM2 is modulated with a phase profile encoding an additional complex amplitude and phase profile onto the incoming beams (48). These five phase profiles are carefully designed to introduce a desired correlation between five spectral subregions with multiple $LG_{\ell,p}$ modes as described in Eq. (2). The reflected beams from SLM2 are then imaged to a single spot in the imaging plane. Another SLM3 at the imaging plane is modulated with a superposition of grating patterns that combine all the modes and frequencies into one STOB. A quarter-wave plate is then used to add SAM to the synthesized STOBs. A third 4*f* system images the STOBs to a semiconductor camera for detection. A limited aperture is placed in the Fourier plane to filter unwanted orders from the grating pattern on SLM3. More experimental details are described in Supplementary Note 5.

**Reconstruction of the spatiotemporal profiles**

To characterize the spatiotemporal properties of the synthesized STOBs, we use an off-axis interference measurement (49) to resolve the spatial, temporal and polarization profiles (Fig. S1 in Supplementary Note 5). The reference copy of the short-pulsed laser is transmitted to a motorized delay stage, which is swept within a range of 1.6 mm with a step size of 20 μm, providing a temporal range and resolution of 10.67 ps and 133 fs, respectively. The temporal resolution might be degraded due to the pulse width of around 299 fs. A pair of rotatable linear polarizers combined with a variable neutral density filter adjusts the intensity of the delayed pulse. To ensure the temporal coherence between the synthesized STOB and the reference pulse, an additional fixed delay of around 2 m is introduced by reflecting the pulse through three mirrors. To further ensure that the reference pulse has a much larger beam size than that of the synthesized STOB, a pair of lenses ($f_1$ = −100 mm and $f_2$ = 300 mm) is used to magnify the reference beam by about three times. Finally, a beam splitter combines the reference pulse with the synthesized STOB with a tilt angle, forming an off-axis interference pattern. The time-averaged intensity profiles of this pattern are recorded by a camera for the reconstruction of the synthesized STOB.

The intensity profile recorded by the camera at a specific delay $t_0$ at a distance $z_0$ is given by $I(x,y,z_0,t_0) = \int \left|E_{STOB}(x,y,z_0,t) + E_{ref}(x,y,z_0,t)\right|^2 dt$. Under the conditions that: (i) the reference pulse has a much larger size than that of the synthesized STOB and can be approximated as a plane wave with a tilt angle along the *x*-axis; (ii) both the synthesized STOB and the reference pulse are on the same polarization; and (iii) the reference pulse width (~ 299 fs) is much shorter than the STOB pulse width (~ 5 ps), the intensity profile can be written as (25):

$$I(x,y,z_0,t_0) \approx A(x,y,z_0) + A_{ref} \cdot e^{-jk_{ref}(x)\cdot x} \cdot E^*_{STOB}(x,y,z_0,t_0) + A_{ref} \cdot e^{jk_{ref}(x)\cdot x} \cdot E_{STOB}(x,y,z_0,t_0), \quad (6)$$

where $A(x,y,z_0)$ is a delay-independent term contributed from both the electric fields $E_{STOB}$ and $E_{ref}$, $A_{ref}$ is a constant coefficient contributed from the approximately plane-wave $E_{ref}$, and $k_{ref}(x)$ is the *x* component of the wavevector of $E_{ref}$. If $k_{ref}(x)$ is large enough that all three terms in Eq. (6) can be separated in the Fourier transformation of $I(x,y,z_0,t_0)$, we can filter out the third term to retrieve the STOB's electric field $E_{STOB}$ as

$$E_{STOB}(x,y,z_0,t_0) \approx e^{-jk_{ref}(x)\cdot x} \cdot \mathcal{F}^{-1}\{\mathcal{F}[I(x,y,z_0,t_0)] \cdot f(k_x,k_y)\}, \quad (7)$$

where $\mathcal{F}$ and $\mathcal{F}^{-1}$ are the Fourier and inverse Fourier transformation operators, respectively, and $f(k_x,k_y)$ is a Fourier filter that selects the third term in Eq. (6).

To further reconstruct the STOB polarization for each position on the *x*-*y* plane at a specific delay $t_0$ at a distance $z_0$, we use a quarter-wave plate and a polarizer to generate a series of combinations of the *x* and *y* components of $E_{STOB}$. These combinations of orthogonal polarizations can be expressed in a form as $E'_{STOB}(x,y,z_0,t_0) = a \cdot E_x(x,y,z_0,t_0) + b \cdot E_y(x,y,z_0,t_0)$, where $\hat{\mathbf{e}} = (E_x, E_y)$ represents the STOB polarization, *a*nd a and b are variable coefficients. Finally, we can use our reference approach as described in Eq. (7) to retrieve a series of values $E'_{STOB}(x, y, z_0, t_0)$, and the quarter-wave-plate method (58) to reconstruct the polarization of the synthesized STOBs.

**Generalization for the synthesis of STOBs exhibiting AM precession**

In the Results section, we demonstrate the experimental synthesis of STOBs with limited precession dynamics as illustrative examples of STOBs exhibiting SAM and OAM precessions (Figs. 5a to 5c). Here we generalize the approach to STOBs with a broader range of precession dynamics, such as higher OAM values, larger precession angles, and higher precession frequencies (Figs. 5g to 5i). We synthesize the STOB exhibiting AM precession in two steps: first, a gradient phase profile is added to a conventional $LG_{\bar{\ell},\bar{p}}$ beam to obtain the electric field of a tilted beam as $LG_{\bar{\ell},\bar{p}}(x, y, z) \cdot e^{-jk_p x} \cdot s(t) \cdot e^{j\omega_0 t} \cdot \hat{\mathbf{e}}$, and a dynamic precession around the *z*-axis is subsequently introduced to the propagation direction. Here, $LG_{\bar{\ell},\bar{p}}(x, y, z)$ represents the complex amplitude and phase profiles of the conventional $LG_{\bar{\ell},\bar{p}}$ beam, $\mathrm{e}^{-jk_p x}$ is the initial phase gradient along the *x*-axis, $s(t)$ is the $LG_{\bar{\ell},\bar{p}}$ beam temporal pulse shape (or envelope), $e^{j\omega_0 t}$ is the $LG_{\bar{\ell},\bar{p}}$ beam temporal phase with $\omega_0$ as the center angular frequency, and $\hat{\mathbf{e}} = (E_x, E_y)$ is the polarization vector.

Based on the modal decomposition method (46, 47), a tilted $LG_{\bar{\ell},\bar{p}}$ beam with arbitrary $(\bar{\ell}, \bar{p})$ values can be expressed as a combination of multiple modes in a complete modal basis set, such as Hermite–Gaussian, Bessel–Gaussian, and $LG$ modal basis sets (3). For instance, the tilted $LG_{\bar{\ell},\bar{p}}$ beam at a distance $z$ can be expressed in $LG$ modal basis sets as $\mathbf{E}_1(x, y, z, t) = \sum\sum_{\ell,p} C_{\ell,p} LG_{\ell,p}(x, y, z) \cdot s(t) \cdot e^{j\omega_0 t} \cdot \hat{\mathbf{e}}$. For synthesizing STOBs exhibiting AM precession in our work, the *y* polarization component of the STOB electric field has either a zero value or the same spatial amplitude and phase profiles as those of the *x* polarization component. Therefore, the coefficient $C_{\ell,p}$ for both polarization components can be calculated in a form as $C_{\ell,p} = \iint LG_{\bar{\ell},\bar{p}}(x, y, z) \cdot e^{-jk_p x} \cdot \left[LG_{\ell,p}(x, y, z)\right]^* dxdy$, where $k_p$ is the wavevector of the gradient phase, and $*$ denotes complex conjugation. The precession of the tilted $LG_{\bar{\ell},\bar{p}}$ beam around the *z*-axis can be understood as a rotation motion of its amplitude and phase profiles around the *z*-axis. Such a rotation motion introduces an angular frequency shift of $\ell\omega_p$ to each $LG_{\ell,p}$ mode, with its center angular frequency $\omega_0$ shifted to $\omega_0 + \ell\omega_p$. Therefore, the electric field $\mathbf{E}_1(x, y, z, t)$ on a single center frequency component carrying multiple $LG_{\ell,p}$ modes is split into different frequency components in a form as $\mathbf{E}_{STOB}(x, y, z, t) = \sum\sum_{\ell,p} C_{\ell,p} LG_{\ell,p}(x, y, z) \cdot s(t) \cdot e^{j(\omega_0 + \ell\omega_p)t} \cdot \hat{\mathbf{e}}$ (Eq. (2)).

The expression of $\mathbf{E}_{STOB}(x, y, z, t)$ indicates that an STOB exhibiting AM precession carrying an arbitrary OAM value of $\bar{\ell}$ can be synthesized by combining multiple frequency components, each carrying the same temporal envelope but multiple $LG_{\ell,p}$ modes. Additionally, this approach can be generalized to synthesize an AM-precession $LG_{\bar{\ell},\bar{p}}$ beam with arbitrary $(\bar{\ell}, \bar{p})$ values via judicious selection of the coefficient $C_{\ell,p}$ for each $LG_{\ell,p}$ mode (24, 35, 36). We note that our experimental illustrative examples of AM-precession STOBs carry a small OAM value and exhibit limited precession dynamics. Extending these ranges spreads the STOBs' power over many more modes and frequency components. However, we believe that the synthesis of STOBs exhibiting AM precession carrying high OAM values (*e.g.*, $\bar{\ell} = 100$) with large precession angles (*e.g.*, 100 mrad) could be achieved using other approaches for the correlation of multiple DoFs, such as multi-plane light converters (52, 53), diffractive deep neural networks (59, 60) and integrated micro-rings (38, 39) as discussed in the Discussion section.

**AM vector of a time slice**

The AM vectors are computed for every time slice and each of the 100 repetitions of a data set, and then averaged over the repetitions (Supplementary Note 4). The transverse field of a slice, $\mathbf{E}_{\perp} = E\,\hat{\mathbf{e}}$, is the reconstructed field $E(x, y)$ multiplied by a uniform Jones vector: $\hat{\mathbf{e}} = \hat{\mathbf{x}}$ for linear polarization and, for circular polarization, the Jones vector with the helicity $s = s_3/s_0$ measured for that slice. With $k_z = \left(k_0^2 - k_x^2 - k_y^2\right)^{1/2}$, the longitudinal field and the magnetic field follow from the free-space relations

$$E_z = -\frac{k_x E_x + k_y E_y}{k_z}, \qquad \mathbf{H} = \frac{\mathbf{k}\times\mathbf{E}}{\mu_0 \omega_0}. \tag{8}$$

From the complete fields we compute the dual-symmetric energy density $w = (\varepsilon_0|\mathbf{E}|^2 + \mu_0|\mathbf{H}|^2)/4$, the canonical momentum density $\mathbf{p}^o = \mathrm{Im}[\varepsilon_0\mathbf{E}^* \cdot (\nabla)\mathbf{E} + \mu_0\mathbf{H}^* \cdot (\nabla)\mathbf{H}]/4\omega_0$, the spin density $\mathbf{s} = \mathrm{Im}[\varepsilon_0\mathbf{E}^* \times \mathbf{E} + \mu_0\mathbf{H}^* \times \mathbf{H}]/4\omega_0$ and the Poynting momentum density $\mathbf{p}^P = \mathrm{Re}(\mathbf{E}^* \times \mathbf{H})/2c^2$ (61, 62), with the transverse gradients taken in the spectral domain and $\partial_z \to ik_z$ for the longitudinal ones. The number of photons per unit length of the slice is $N = \int w\, dA/\hbar\omega_0$ and its energy centroid is $\mathbf{r}_c = \int \mathbf{r}\, w\, dA/\int w\, dA$. The AM vectors per photon, in units of $\hbar$, are

$$\mathbf{L} = \frac{1}{\hbar N}\int (\mathbf{r} - \mathbf{r}_c) \times \mathbf{p}^o dA, \quad \mathbf{S} = \frac{1}{\hbar N}\int \mathbf{s}\, dA, \quad \mathbf{J} = \frac{1}{\hbar N}\int (\mathbf{r} - \mathbf{r}_c) \times \mathbf{p}^P dA, \tag{9}$$

with the lever arm $\mathbf{r} = (x, y, 0)$ in the plane of the slice. The mean wavevector, whose direction defines the measured tilt $\boldsymbol{\alpha} = \mathbf{P}_{\perp}/P_z$, is $\mathbf{P} = \int \mathbf{p}^o dA/\hbar N$. See Supplementary Note 4 for more details.

**Acknowledgments:**

**Funding:** This work was supported by the National Natural Science Foundation of China (NSFC) (62125503, 62261160388), the Natural Science Foundation of Hubei Province of China (2023AFA028), and the Hubei Optical Fundamental Research Center (HBO2025TQ004). Z.Z. acknowledges funding support through a Humboldt Research Fellowship from the Alexander von Humboldt Foundation.

**Author contributions:**

Conceptualization: ZZ, JW

Methodology: ZZ, YL, XL, JW

Investigation: ZZ, JL, YL

Visualization: ZZ, YL

Funding acquisition: ZZ, JW

Experimental conditions: JW

Project administration: JW

Supervision: JW

Writing – original draft: ZZ, JL, YL, XL, JW

**Competing interests:** The authors declare no competing interests.

**Data, materials and code availability:** All data, theory details, simulation details that support the findings of this study are available from the corresponding authors on reasonable request.

**Supplementary Materials**

Materials and Methods

Supplementary Text

Figs. S1 to S9

Supplementary Movie 1 to 5

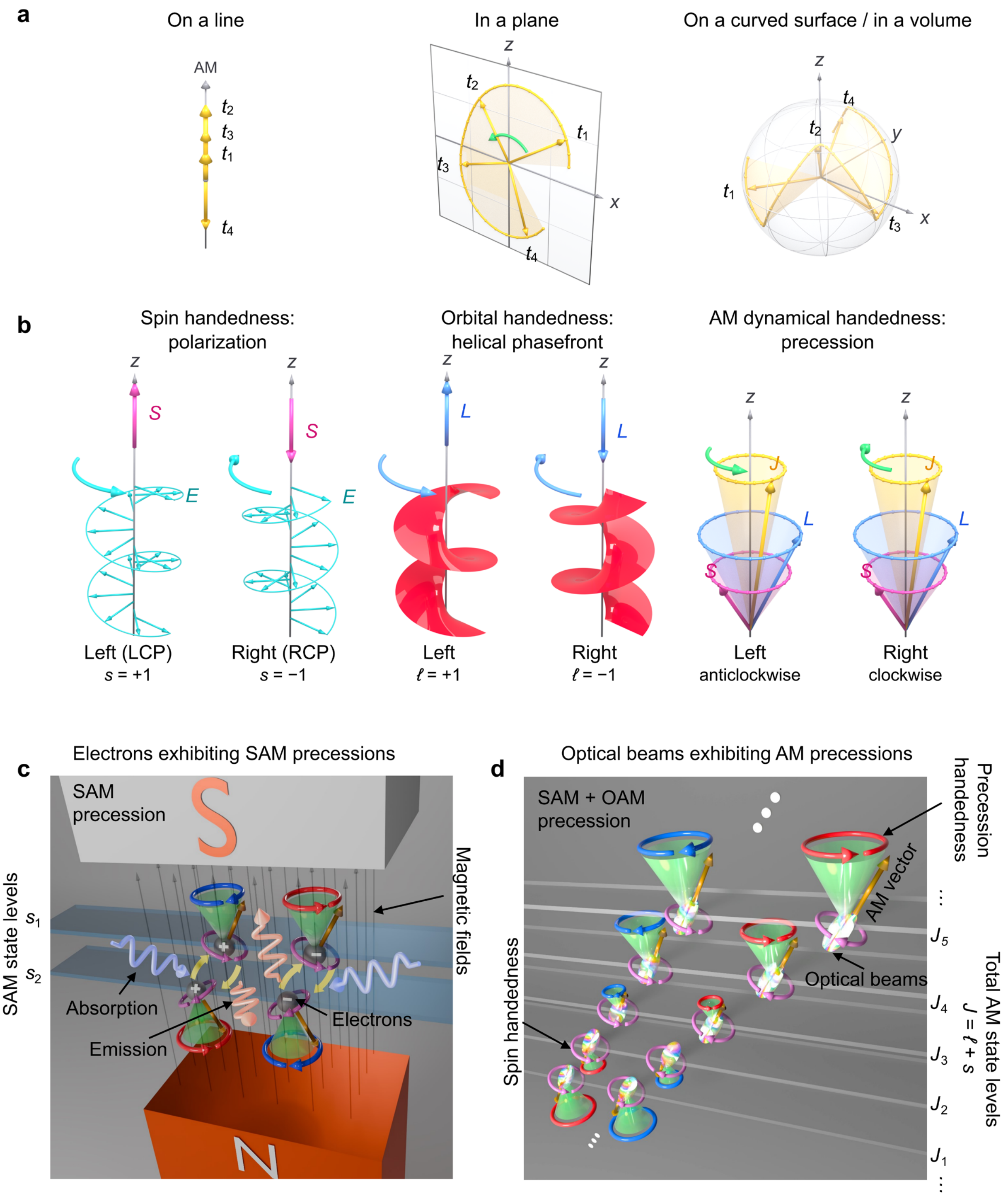

**Fig. 1. Illustrations of spatiotemporal control of 3D spin and orbital angular momenta vectors for the precession of optical beams. (a)** Comparison of the dynamic behaviors of angular momentum (AM) vectors on a line, in a plane and in a volume. **(b)** Comparison of the spin, orbital and AM dynamic handedness of optical beams. (**c**) Electrons exhibiting SAM precession driven by the torque that an external magnetic field exerts on their magnetic moment, and (d) STOBs exhibiting both SAM and OAM precession set by the frequency–OAM correlation, with no external torque. The SAM of charged particles in materials points in random directions in the absence of an external magnetic field. In an external magnetic field $\boldsymbol{B}$, the SAM precesses about

the direction of $\boldsymbol{B}$ at a fixed precession angle, which is known as Larmor precession. Additionally, as illustrated in (**c**), the SAM precession of electrons can be changed: (i) the precession handedness can be flipped between anticlockwise (red circular arrows) and clockwise (blue circular arrows) when the sign of the charge, and hence of the gyromagnetic ratio, is reversed, and (ii) electrons exhibiting SAM precession can flip the SAM sign by absorbing (light blue arrows) and emitting (light pink arrows) a radio-frequency (RF) wave at the resonance frequency. Analogously, conventional $LG_{\ell,p}$ optical pulses propagating along the $z$-axis can carry an OAM either parallel or antiparallel to the $z$-axis. Furthermore, STOBs exhibiting both SAM and OAM precessions in (**d**) can be synthesized with programmable dynamics by correlating frequency and OAM. LCP (or RCP): left-hand (or right-hand) circular polarization; SAM: spin angular momentum; OAM: orbital angular momentum.

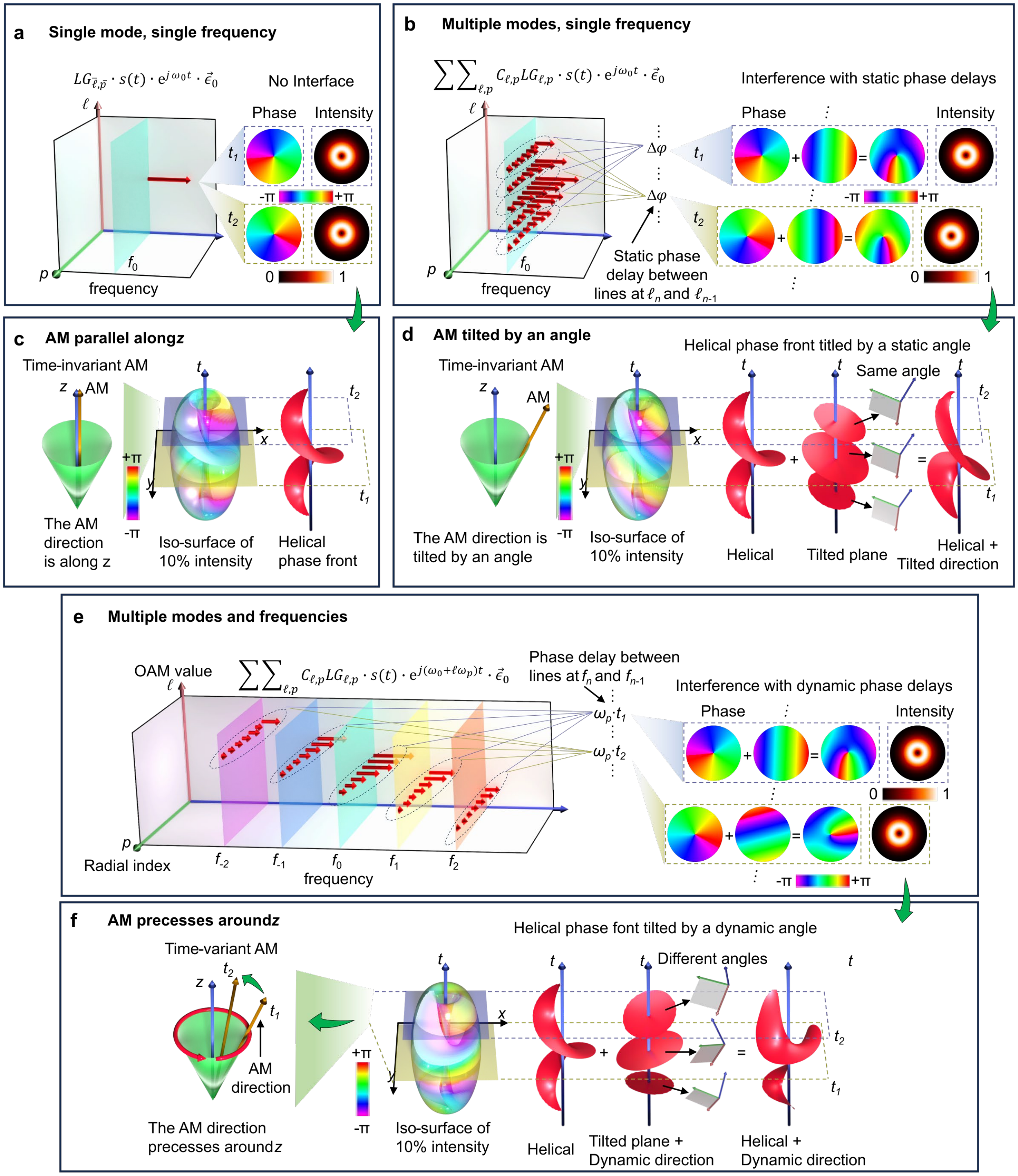

**Fig. 2. Concept for the synthesis of STOBs exhibiting AM precession via frequency–OAM correlation.** (**a**), (**c**) A conventional $LG_{1,0}$ beam on a single frequency component has a ring-like intensity profile and a 'twisting' phase profile rotating around its beam center over time. It produces an iso-intensity surface (10% of the maximum intensity) in a 'vortex hollow' cylindrical shape and an iso-phase surface (a zero-phase value) in a helical shape in the ($x$, $y$, $t$) coordinates. The beam has an OAM with a time-invariant direction along the positive z-axis. (**b**), (**d**) The interferogram of multiple $LG_{\ell,p}$ modes containing multiple $\ell$ values and $p$ values carried by a single frequency component. The interference with static phase delays ($\Delta\varphi$) produces a ring-like intensity

profile and a ‘fork’ phase profile. Therefore, its helical iso-phase surface is tilted by adding tilted iso-phase planes to the initial helical surface in (**d**). Its OAM also has a time-invariant direction, tilted from the z-axis. (**e**), (**f**) The combination of multiple frequency components, each carrying a different $\ell$ value and multiple $p$ values. The interference with dynamic phase delays ($\Delta\varphi = \Delta\omega t$) produces a ‘fork’ phase profile rotating around the beam center over time. Therefore, its helical iso-phase surface is tilted by an angle with a time-variant value, leading to the dynamic AM precession in (**f**).

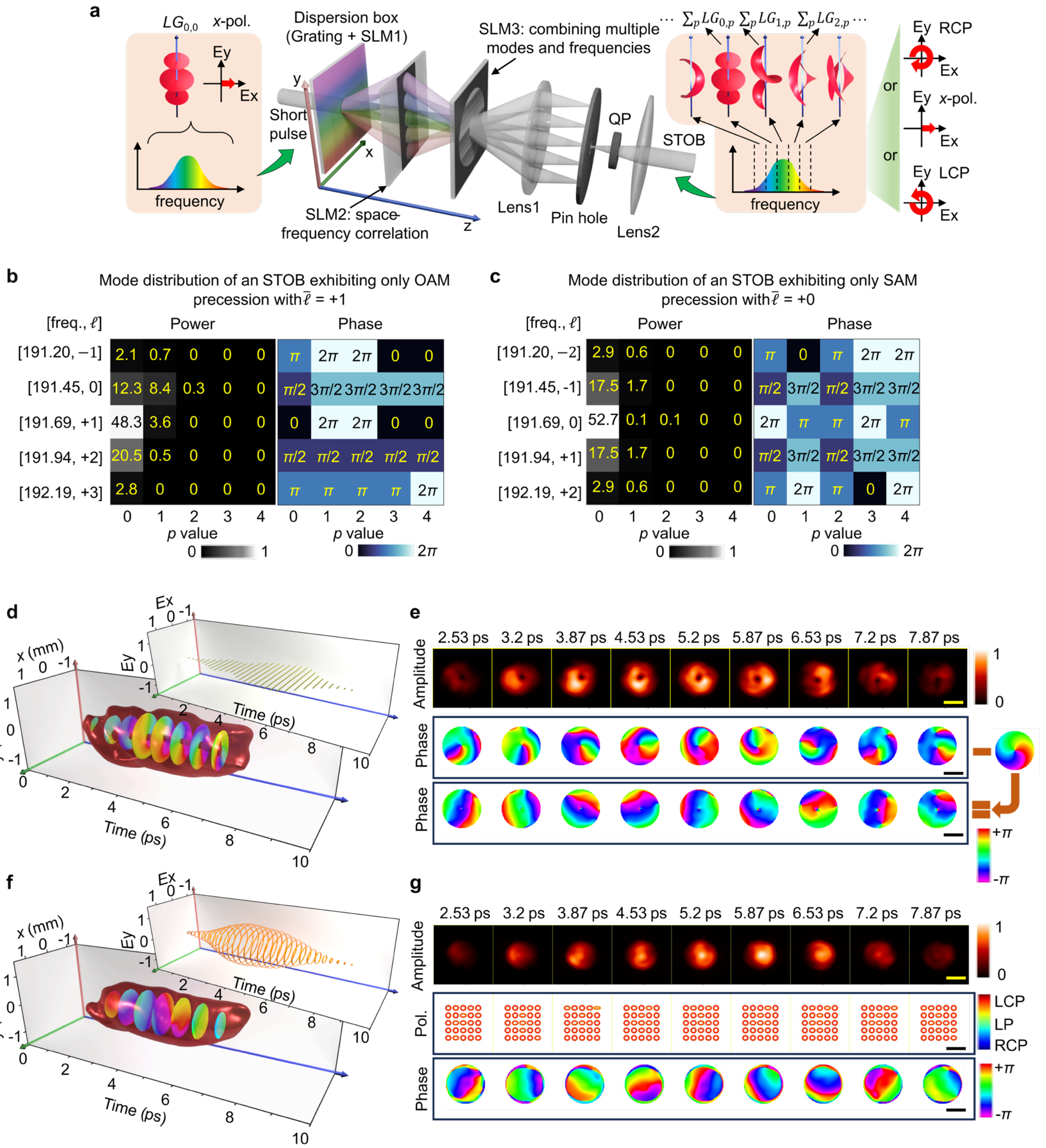

**Fig. 3. Experimental synthesis of STOBs exhibiting only OAM and only SAM precessions.** (**a**) Conceptual scheme for the synthesis of STOBs by correlating multiple degrees of freedom (DoFs). The process begins with a short-pulsed laser. A dispersion box, consisting of a blazed grating, a 4f imaging system, and an SLM1, selects five equally spaced spectral subregions and maps them onto five areas along the $y$-axis on a second SLM2. We show the dispersion box using one simple cube for clarity (see Fig. S1 in Supplementary Note 5 for the actual setup). Each area on SLM2 is modulated with a complex spatial amplitude and phase profile to tailor the correlation between spatial modes and frequencies. A third SLM3, with a superposition of multiple grating patterns, combines all the modes and frequencies into one STOB exhibiting AM precessions. The beam path is shown along the $z$-axis for clarity, although it is folded in the experimental setup (Fig. S1 in Supplementary Note 5). (**b**) The mode distribution of an STOB exhibiting only OAM

precession with $\overline{\ell} = +1$. (**c**) The mode distribution of an STOB exhibiting only SAM precession with $s = +1$ and $\overline{\ell} = 0$. Frequency unit in THz. (**d**) The reconstructed spatial amplitude/intensity, phase, and polarization profiles of an STOB exhibiting only OAM precession with $\overline{\ell} = +1$. The iso-intensity surface (10% of the maximum intensity, shown as a red shell in the left-bottom coordinate) has a 'vortex hollow' cylindrical shape. The inserted colored slices represent the phase profiles in the *x*-*y* plane at different times. The STOB's polarization at (0, 0.25 mm) in the *x*-*y* plane remains along the *x*-axis, indicating a zero SAM value (the right-top coordinate). (**e**) The 2D temporal evolutions of the STOB's amplitude and phase profiles. The STOB has a phase profile that combines a 'twisting' phase and a gradient phase 'rotating' around the beam center. (**f**) The reconstructed iso-intensity surface and (**g**) the corresponding 2D temporal evolutions of an STOB exhibiting only SAM precession with $s = +1$. The synthesized STOB has a dot-like amplitude shape (first row), a polarization of almost LCP ($s \approx +1$, second row), and a gradient phase with a dynamically changing slope direction over time. SLM: spatial light modulator; QP: quarter-wave plate; Pol.: polarization; LP: linear polarization; freq.: frequency (THz). Color bar in (**e**) and (**g**): 1 mm.

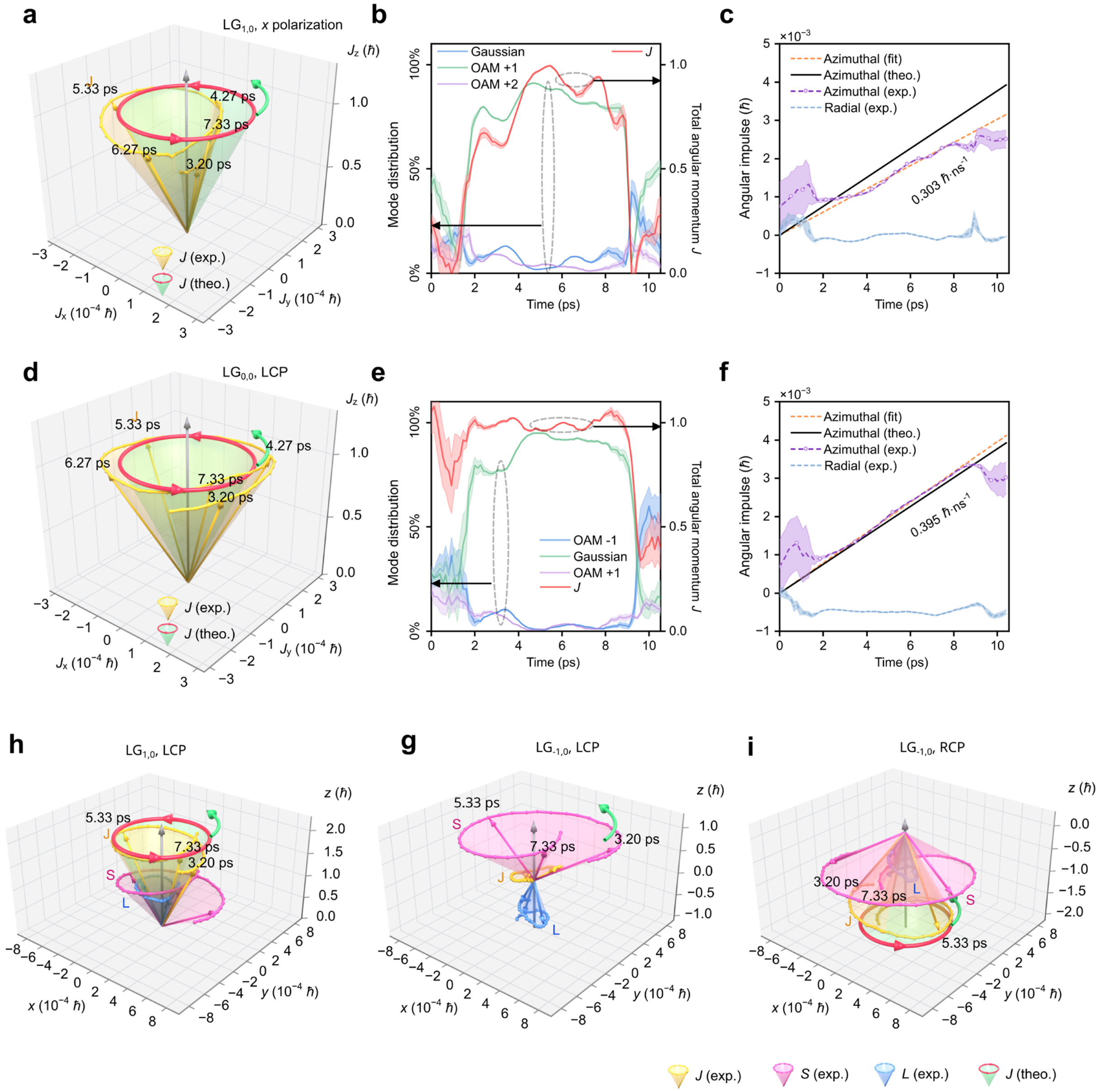

**Fig. 4. Characterizing the AM precession, AM value, and angular impulse.** (**a**) An OAM-only STOB exhibiting precession of the total AM **J** around the beam center with a period of around 4.13 ps. (**b**) The corresponding mode distribution and the total AM value |**J**| of the synthesized STOB. (**c**) The synthesized STOB has an angular impulse due to its dynamic AM change, with the azimuthal component appearing linear dependent on time and the radial component and the radial component fluctuating around zero. (**d**), (e) and (**f**) illustrate the same characterizations for an SAM-only STOB with $s = +1$ exhibiting AM precession. Similarly, although the synthesized STOB has a near-zero OAM value, its AM **J** still precesses around its beam center driven by its SAM. (**g**), **(h) and** (**i**) illustrate the anticlockwise precession of OAM **L**, SAM **S** and total AM **J** for synthesized STOBs carrying OAM and SAM of $(\bar{\ell}, s) = (+1, +1)$, $(-1, +1)$ and $(-1, -1)$, respectively. The spin and orbital handedness of the STOBs can be changed while the handedness of their AM precession stays the same. Exp.: experiment; Theo.: theory.

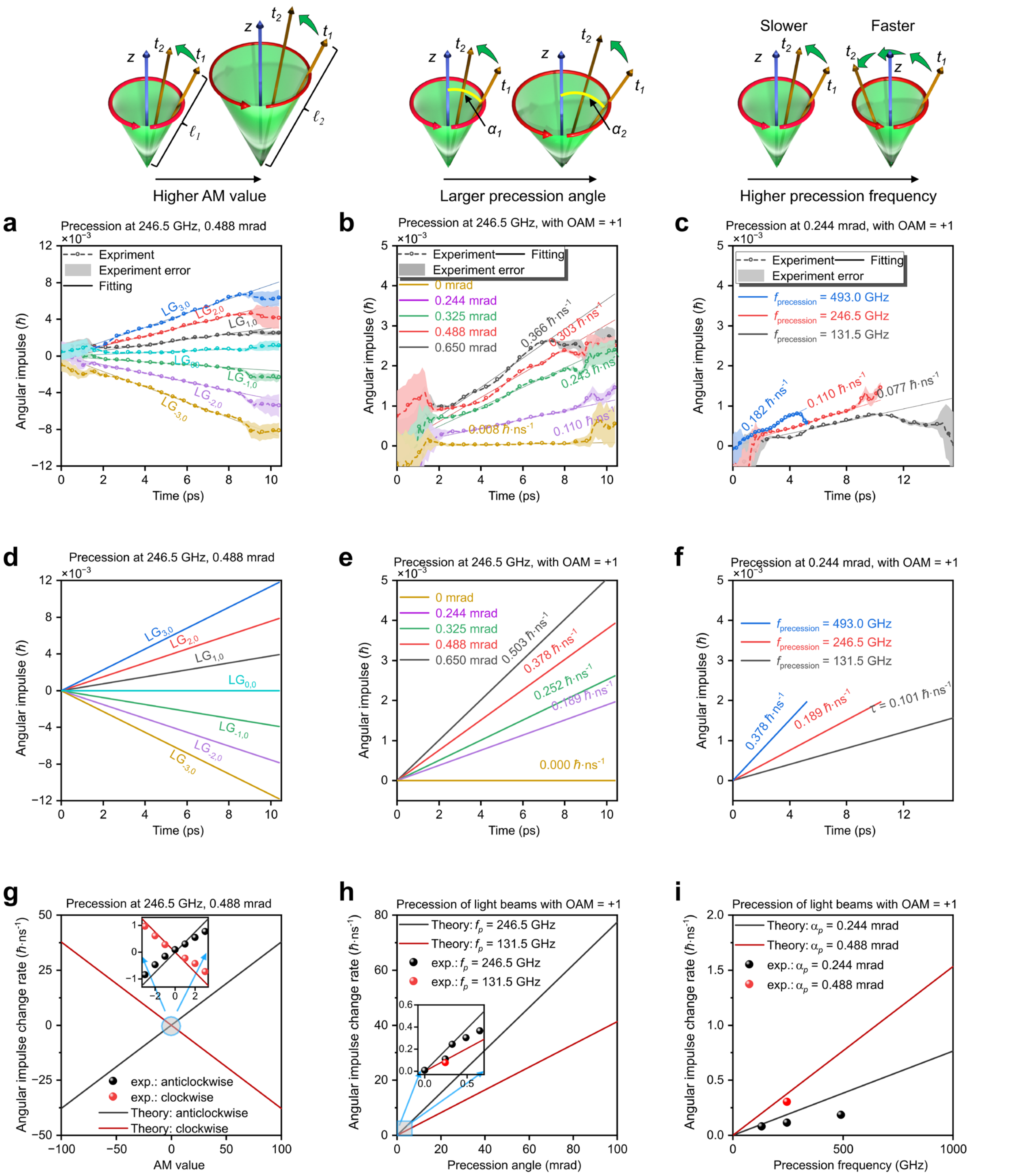

**Fig. 5. Programmable AM precession via the frequency–OAM correlation.** The AM precessions can be controlled by selecting the correlation between the STOB's frequency and OAM components. We first show experimentally that the synthesized STOB has a controllable (**a**) total AM value, (**b**) precession angle, and (**c**) precession speed. Experimental errors smaller than $5 \times 10^{-4}\ \hbar$ are barely visible at the scale of (a)–(c). (**d**), (**e**), and (**f**) indicate the theoretical values calculated using Eqs. (3) and (5), showing the same trends as the experimental results. The limited dynamic ranges could potentially be extended. (**g**), (**h**), and (**i**) show the predicted change rate of the angular impulse versus the AM value, precession angle, and precession frequency in broader ranges. The measurements of (a)–(**c**) are included to partially verify the prediction accuracy.